\documentclass[a4paper,11pt]{article}
\usepackage{jcappub} 

\RequirePackage{graphicx}
\usepackage{subfigure}

\title{\boldmath Cosmological constraints on dark energy models using DESI BAO 2024}

\author[a]{Jie Zheng,}
\author[a]{Da-chun Qiang,}
\author[a]{and Zhi-Qiang You}
\affiliation[a]{Institute for Gravitational Wave Astronomy, Henan Academy of Sciences,\\No.228 Chongshili, Zhengzhou, China}

\emailAdd{zhengjie@mail.bnu.edu.cn}

\abstract{Recently, the measurements of baryon acoustic oscillations (BAO) by the Dark Energy Spectroscopic Instrument (DESI) indicate a potential deviation from the standard $\Lambda$CDM model. 
Some studies suggest that the data points from the luminous red galaxies (LRG) survey in DESI BAO data may contribute to this discrepancy. 
In this work, our main goal is to investigate whether this deviation is caused by the parameterization of the equation of state (EoS) of dark energy (DE).
Hence, we have examined four popular parameterized dark energy models in our analysis: the Chevallier-Polarski-Linder (CPL), Barboza-Alcaniz (BA), Jassal-Bagla-Padmanabhan (JBP), and Feng-Shen-Li-Li (FSLL) parameterizations.
Considering that LRG1 and LRG2 data points may lead to deviation from the $\Lambda$CDM model, we use two versions of DESI BAO data, differing in whether these data points are included.
Additionally, to break the parameter degeneracies and obtain robust constraint results, we introduce Type Ia supernovae (SNe Ia) and quasars (QSO) in our analysis.
Our findings indicate that in these parameterizations, the deviation from ($w_0$,$w_{1}$)=(-1,0) becomes more pronounced when using the combined data from DESI BAO, SNe Ia, and QSO compilations. Here, $w_{0}$ and $w_{1}$ represent the EoS of DE.
It suggests that the parameterizations of the EoS of DE have little impact on the deviation from the $\Lambda$CDM model. 
Besides, our analysis potentially hints that dark energy may have dynamic properties.
In addition, the results obtained from different BAO datasets demonstrate that the LRG1 and LRG2 data points do indeed contribute to a deviation from the $\Lambda$CDM model. 
Finally, according to the statistical criteria, the Akaike Information Criterion (AIC) and the Bayesian Information Criterion (BIC), the joint constraints provide substantial observational support to the BA and FSLL models.}

\begin{document}
\maketitle
\flushbottom

\section{Introduction}
\label{sec:intro}

Various observations have provided a crucial cosmological discovery that the universe is undergoing an accelerated expansion \cite{1998AJ....116.1009R,1999ApJ...517..565P}. 
Despite the considerable precision of current observational data, understanding the mechanisms responsible for this late-time acceleration remains challenging.
It is widely hypothesized that the concept of dark energy (DE), a cosmic fluid with negative pressure, is responsible for this phenomenon. 
The standard model of cosmology, often referred to as the $\Lambda$ CDM model, is also the simplest dark energy model. In this model, the universe is composed of cold dark matter, baryons, negligible radiation, and dark energy. The cosmological constant $\Lambda$, interpreted as the energy of the quantum vacuum, corresponds to the equation of state (EoS) of DE $w=-1$ \cite{2001LRR.....4....1C}. 
The $\Lambda$CDM model has been remarkably successful in explaining various cosmological observations, such as the cosmic microwave background (CMB) \cite{planck2018}, Type Ia supernovae (SNe Ia) \cite{pantheon,pantheonp} and the strong gravitational lensing (SGL) \cite{2017MNRAS.465.4895W,2020MNRAS.498.1420W}. Nevertheless, it is struggled by several theoretical problems, including the well-known fine-tuning and coincidence problems \cite{weinberg1,weinberg2}, along with the Hubble tension problem \cite{2019NatAs...3..891V,2021CQGra..38o3001D}. 
As a result, the dynamical DE models with the parameterizations of the EoS of DE have been proposed and investigated.

The Baryon Acoustic Oscillation (BAO) is a crucial feature in the large-scale distribution of matter in the universe, arising from the interaction between photons and baryons in the very early universe. This characteristic is imprinted on the matter distribution of the early universe by physics around recombination and earlier. With the expansion of the universe, these oscillations are stretched, appearing at a comoving galaxy separation of $r_{d} \sim 150$Mpc. Consequently, BAO serves as a standard ruler, its length determined by the physics of the early universe, providing a useful way to measure the EoS of DE, spatial curvature, and the expansion history of the universe \cite{2013PhR...530...87W,2021PhRvD.103h3533A}.
In 2005, the BAO feature was first detected by the Sloan Digital Sky Survey (SDSS) \cite{eisenstein20005} and the Anglo-Australian Telescope Two-degree Field Galaxy Redshift Survey (2dF) \cite{2005MNRAS.362..505C}. Subsequent measurements improved precision using various surveys, including the SDSS-III Luminous Red Galaxy Sample \cite{2007MNRAS.381.1053P,2010MNRAS.401.2148P}, the WiggleZ Dark Energy Survey \cite{2011MNRAS.415.2892B,2011MNRAS.418.1707B,2014MNRAS.441.3524K}, and the 6-degree Field Galaxy Survey \cite{2011MNRAS.416.3017B,2018MNRAS.481.2371C}. Extensions of the SDSS resulted in more precise percent-level BAO measurements using the Baryon Oscillation Spectroscopic Survey (BOSS) \cite{2012MNRAS.427.3435A,2014MNRAS.441...24A,2017MNRAS.470.2617A} at $z<0.7$ and its extension, eBOSS \cite{2018MNRAS.473.4773A,2021MNRAS.500..736B,2021MNRAS.500.1201H} at $z>0.7$.
BAOs in the Lyman-$\alpha$ forest from quasar surveys extended measurements to $z>2$ \cite{2013A&A...552A..96B,2014JCAP...05..027F,2015A&A...574A..59D,2017A&A...603A..12B}. Combining the results obtained from these various surveys, the BAO measurements effectively mapped out the relationship between cosmological distance and redshift with an impressive accuracy of 1\% to 2\% for $z<2.5$ \cite{2014JCAP...05..027F,2015PhRvD..92l3516A}.
Recently, the Dark Energy Spectroscopic Instrument (DESI) collaboration released its first year of data (DR1) \cite{2024arXiv240403000D,2024arXiv240403001D}, which covers six different types of tracers to trace the distribution of neutral hydrogen, including the bright galaxy survey (BGS), luminous red galaxies (LRG), emission line galaxies (ELG), quasars, and Lyman-$\alpha$ (Ly$\alpha$) forest quasars. 
The corresponding cosmological analysis of DESI BAO data, detailed in Ref.~\cite{2024arXiv240403002D}, found no evidence for $w\neq -1$ under the assumption of a flat $w$CDM model. 
However, in the $w_{0}w_{a}$CDM model, where the EoS of DE varies with time as $w(a)=w_{0}+(1-a)w_{a}$, $a$ is the scale factor, the result obtained from DESI BAO and CMB data indicates a significant tension with the $\Lambda$CDM model. This tension increases to a 3.9$\sigma$ level when combined with DESY5 SNe Ia data \cite{2024arXiv240403002D}.

The deviations from the $\Lambda$CDM model observed in the DESI BAO data have sparked extensive discussions on dark energy. Numerous studies have delved into the reasons behind these discrepancies, such as inconsistencies between the
different datasets. For instance, \cite{2024arXiv240502168W} demonstrates that excluding the LRG1($z_{\text{eff}} = 0.51$) and LRG2 ($z_{\text{eff}} = 0.71$) data individually results in $w = -1$ falling within the 2$\sigma$ contour. \cite{2024arXiv240704385L} also points out that the deviation from the $\Lambda$CDM model is predominantly driven by the LRG1 and LRG2 data.
Moreover, \cite{2024arXiv240804432G} finds that the BAO measurements from the Sloan Digital Sky Survey (SDSS) and the DESI collaboration are inconsistent with each other. This inconsistency may lead to biased and divergent results regarding the validity of the standard model or the suggestion of new physics.

In this work, we investigate whether this deviation arises from a specific parameterization. Hence, we consider more comprehensive dynamical dark energy parameterizations, including the Jassal–Bagla–Padmanabhan (JBP) parameterization \cite{2005MNRAS.356L..11J,2021Univ....7..163M}, the Barboza-Alcaniz (BA) parameterization \cite{2008PhLB..666..415B,2022JCAP...03..060E}, and the Feng–Shen–Li–Li (FSLL) parameterization \cite{2012JCAP...09..023F}. In addition, the combination of the different observations would effectively break the degeneracy between cosmological parameters \cite{Lian2021,zheng_EPJC}. Therefore, we introduce 2421 optically selected QSO data with spectroscopic redshift up to $z\sim 7$ \cite{2020A&A...642A.150L}, and the newly SNe Ia data, ``Pantheon+" sample \cite{pantheonp}, into our analysis. 

The remainder of the paper is organized as follows. We introduce the data and the analysis method used in this work in Sec.~\ref{sec:2}. In Sec.~\ref{sec:3}, we summarize the cosmological models to be analyzed. We report the constraint results in Sec.\ref{sec:4}. Finally, we summarize our conclusions in Sec.\ref{sec:5}.

\section{Observational Data}
\label{sec:2}

\subsection{The DESI measurements of the baryonic acoustic oscillations}

\begin{table}[htbp]
\centering
\begin{tabular}{l|c|c|c|c}
\hline
Tracer & $z_{\mathrm{eff}}$ & $d_{M}/r_{d}$ & $d_{H}/r_{d}$ & $d_{V}/r_{d}$\\
\hline
BGS & 0.295 & - & - & $7.93\pm0.15$\\
LRG & 0.510 & $13.62\pm0.25 $& $20.98\pm0.61$ & -\\
LRG & 0.706 & $16.85\pm0.32$ & $20.08\pm0.60$ & -\\
LRG+ELG & 0.930 & $21.71\pm0.28 $& $17.88\pm0.35 $& -\\
ELG & 1.317 & $27.79\pm0.69$ & $13.82\pm0.42$ & - \\
QSO & 1.491 & - & - & $26.07\pm0.67$ \\
Ly$\alpha$ QSO & 2.330 & $39.71\pm0.94$ & $8.52\pm0.17$ & - \\
\hline
\end{tabular}
\caption{The DESI-BAO samples with tracers, effective redshifts $z_{\mathrm{eff}}$, and ratios $d_{M}/r_{d}$, $d_{H}/r_{d}$, and $d_{V}/r_{d}$, which is reproduced from Ref.~\cite{2024arXiv240403002D}. \label{tab:baodata}}
\end{table}

The clustering of matter in the universe reveals a significant scale, known as the sound horizon, which was imprinted during the early universe's baryon drag epoch. This feature, shaped by physics around recombination and earlier epochs, expands with the universe and appears at a comoving galaxy separation of approximately $r_d \sim 150$ Mpc. 
It is known as BAO, which is seen both as wiggles in the power spectrum of galaxies and as a peak in the two-point correlation function \cite{eisenstein1998,2001MNRAS.327.1297P,2005MNRAS.362..505C,eisenstein20005}. As a result, BAO can serve as the standard ruler for studying the expansion history of the universe \cite{2010deot.book..246B,2013PhR...530...87W,2021PhRvD.103h3533A,2024arXiv240403002D}, providing complementary information to that of SNe Ia and CMB.

DESI is carrying out a State IV survey aimed at significantly enhancing cosmological constraints and it is conducting a 5-year survey that covers 1,4200 square degrees in the redshift range $0.1<z<4.2$ over five years, and its spectroscopic sample size is ten times larger than previous surveys.
The survey covers six different classes of tracers, including low redshift galaxies of the BGS ($0.1<z<0.4$), LRG ($0.4<z<0.6$ and $0.6<z<0.8$), ELG ($1.1<z<1.6$), the combined LRG and ELG sample (LRG+ELG, $0.8<z<1.1$), quasars as direct tracers (QSO, $0.8<z<2.1$), and Lyman-$\alpha$ forest quasars (Ly$\alpha$ QSO, $1.77<z<4.16$) to trace the distribution of neutral hydrogen.
It will provide stringent constraints on the cosmological parameters and offer a robust test for modifications to the general theory of relativity \cite{2016RPPh...79d6902K,2016ARNPS..66...95J,2019LRR....22....1I,2021JCAP...11..050A}.
In this paper, we derive constraints on cosmological models using BAO measurements from DESI's first-year data, listed in Tab.~\ref{tab:baodata}.
It should be noted that all measurements are effectively independent from each other \cite{2024arXiv240403002D} and no covariance matrix is introduced here, while the systematics related to these measurements introduce a negligible offset \cite{2021MNRAS.503.3510G,2024arXiv240403002D}. 
Furthermore, we adopt the prior knowledge of $\Omega_{b}h^{2}=0.02218\pm0.00055$ from the Big Bang Nucleosynthesis \cite{2024arXiv240403002D} to break the degeneracy between $H_0$ and $r_{d}$.

The first observable quantities used in the measurements are expressed as the ratios
\begin{equation}
    \frac{d_{M}(z)}{r_{d}}=\frac{D_{L}(z)}{r_{d}(1+z)},
\end{equation}
where $d_{M}(z)$ is the transverse co-moving distance, and the $r_{d}$ denotes the sound horizon at the drag epoch, which depends on the matter and baryon physical energy densities and the effective number of extra-relativistic degrees of freedom. We adopt the method described in Ref.~\cite{eisenstein1998} to compute $r_{d}$. Theoretically, $D_{\mathrm{L}}$ is determined by the redshift $z$ and cosmological parameters $\hat{p}$ in a specific model:
\begin{equation}
    D_{\mathrm{L}}(z,\hat{p})=\frac{c(1+z)}{H_{0}}\int^{z}_{0} \frac{dz^{\prime}}{E(z^{\prime})}.
\end{equation}
The second observable is 
\begin{equation}
    \frac{d_{H}(z)}{r_{d}}=\frac{c}{r_{d}H(z)},
\end{equation}
where $d_{H}(z)$ is the Hubble distance and $H(z)$ is the Hubble parameter.
And the third observable is
\begin{equation}
    \frac{d_{V}(z)}{r_{d}}=\frac{[zd^{2}_{M}(z)d_{H}(z)]^{1/3}}{r_{d}},
\end{equation}
where $d_{V}(z)$ represents the volume-average angular diameter distance.

The $\chi^{2}$ likelihood function of each ratio, $X=d_{M}/r_{d}, d_{H}/r_{d}, d_{V}/r_{d}$ can be written as
\begin{equation}
\chi^{2}_{\mathrm{BAO}}=\sum_{i}^{N} \{[\frac{X_{i}-X_{\mathrm{th}}}{\sigma_{X_{i}}}]^{2}+\ln (2 \pi \sigma^{2}_{X_{i}}) \}.
\end{equation}

\subsection{The quasar sample}

Quasars are the most luminous and persistent energy sources in our universe, with the integrated luminosities of $10^{44-48}$ erg/s over the UV to the X-ray energy range. 
They bear a considerable potential to serve as cosmological probes by using the non-linear relation between their UV ($L_{\mathrm{UV}}$)and X-ray ($L_{\mathrm{X}}$) emission \cite{1979ApJ...234L...9T,1981ApJ...245..357Z,1982ApJ...262L..17A,2018Natur.553..473B}.
Since 2015, \cite{Risaliti2015} used this approach to provide an independent measurement of their distances, turning quasars into standardizable candles and extending the distance modulus–redshift relation of Tpye Ia supernovae to a redshift range that is still poorly explored ($z>2$; \cite{2015ApJ...815...33R}).
So far, the largest broad-line quasar main sample has been assembled by combining seven different samples from the literature and public archives, consisting of $\sim$ 19000 objects \cite{2020A&A...642A.150L}. The former group includes samples at $z\simeq 3.0-3.3$ by \cite{2019A&A...632A.109N}, $4<z<7$ by \cite{2019A&A...631A.120S}, $z>6$ by \cite{2019A&A...630A.118V}, and the XMM-XXL North quasar sample by \cite{2016MNRAS.457..110M}. Then they complemented this collection by including quasars from a cross-match of optical (i.e., Sloan Digital Sky Survey; SDSS) and X-ray public catalogs (i.e., XMM-Newton and Chandra) \cite{2021A&A...655A.109B}. Finally, they added a local subset of AGN with UV (i.e., International Ultraviolet Explorer) data and X-ray archival information. 
Recently, \cite{2020A&A...642A.150L} put forward a new catalog of 2421 optically selected quasars from QSO X-Ray and UV flux measurements spanning the redshift range $0.009<z<7.52$ with accurate estimates of $L_{\mathrm{UV}}-L_{\mathrm{X}}$, which have been used to constrain cosmological models \cite{2022PhRvD.105j3526L,2022arXiv220406118K,2022MNRAS.510.2753K,2024MNRAS.530.4493Z,2024EPJC...84..444L}. 
The application of this data set to explore cosmological research depends on the empirical relationship between the X-ray and UV luminosities of these high redshift quasars proposed
by \cite{1986ApJ...305...83A}, which leads to the Hubble diagram constructed by quasars \cite{2015ApJ...815...33R,2016ApJ...819..154L,2017AN....338..329R,2017FrASS...4...68B,2019NatAs...3..272R}. 

The non-linear relation between the $L_{\mathrm{UV}}$ and $L_{\mathrm{X}}$ luminosities is given by \cite{2015ApJ...815...33R,Risaliti2015}
\begin{equation}
    \log L_{\mathrm{X}}= \beta + \gamma \log L_{\mathrm{UV}},
\end{equation}
from which the luminosity distance can be derived as:
\begin{equation}
    \log D_{\mathrm{L}}= \frac{[\log F_{\mathrm{X}}-\beta-\gamma(\log F_{\mathrm{UV}}+27.5)]}{2(\gamma-1)}-\frac{1}{2}\log (4\pi) + 28.5,
\end{equation}
where $F_{\mathrm{X}}$ and $F_{\mathrm{UV}}$ are the flux densities at X-ray and UV wavelengths, respectively, and $F=L/4\pi D^{2}_{\mathrm{L}}$. In this equation, $F_{\mathrm{UV}}$ and $F_{\mathrm{X}}$ are normalized to 27.5 (logarithmic scale) and 28.5 (logarithmic scale), separately. Here, the luminosity distance $D_{\mathrm{L}}$ is in units of centimeters (cm). The slope $\gamma$ of the $F_{\mathrm{X}}-F_{\mathrm{UV}}$ relation and the intercept $\beta$ are free parameters. The intercept $\beta$ of the $L_{\mathrm{UV}}-L_{\mathrm{X}}$ relation is associated with the intercept $\hat{\beta}$ of $F_{\mathrm{X}}-F_{\mathrm{UV}}$ relation by:
\begin{equation}
    \hat{\beta}(z)=2(\gamma-1)\log D_{\mathrm{L}}(z) + (\gamma-1) \log 4\pi + \beta.
\end{equation}

The $\chi^{2}$ likelihood function of this QSO sample is defined as:
\begin{equation}
\chi^{2}_{\mathrm{QSO}}= \sum_{i}^{N}(\frac{(y_i-\psi_i)^2}{s_i^2}-\ln s_i^2),
\end{equation}
where $s_{i}^{2}=dy^{2}_{i}+\gamma^{2}dx_{i}^{2}=\mathrm{exp}(2\ln \delta)$ accounts for uncertainties in both $x_i$ ($\log F_{\mathrm{UV}}$) and $y_i$ ($\log F_{\mathrm{X}}$), and $\delta$ represents the global intrinsic dispersion. 
The variable $\psi$ represents the theoretical X-ray flux $F_{\mathrm{X,th}}$, which depends on the cosmological model, and is given by:
\begin{equation}
    \psi=\log F_{\mathrm{X,th}}=\beta + \gamma(\log F_{\mathrm{UV}}+27.5) + 2(\gamma-1)(\log D_{\mathrm{L,th}}-28.5).
\end{equation}
Recently, \cite{2021MNRAS.507..919L} presented a model-independent method to calibrate this sample using the latest SNe Ia data, estimating $\beta=7.735\pm0.244$, $\gamma=0.649\pm0.007$, and $\delta=0.235\pm0.04$, and we apply their calibrated parameters in our analysis.

\subsection{The Pantheon+ Sample of Type Ia supernovae}

SNe Ia, renowned for their standardized luminosity, have proven to be reliable cosmological probes. In the past two decades, plentiful supernovae surveys have been carried out to detect supernovae across a wide range of redshifts.
These include surveys targeting low-redshift supernovae ($0.01 < z < 0.1$) such as CfA1-CfA4 \cite{Riess:1998dv,Jha:2005jg,Hicken:2009df,Hicken:2012zr}, CSP \cite{Krisciunas:2017yoe}, SOUSA \cite{Brown:2014gqa}, Low-$z$ \cite{}, CNIa0.02 \cite{Chen:2020qnp}, Foundation \cite{Foley:2017zdq}, and LOSS \cite{Stahl:2019xzs}. Additionally, four major surveys have probed the redshift range of $z > 0.1$, namely DES \cite{2019ApJ...874..106B}, SNLS \cite{2014A&A...568A..22B}, SDSS \cite{2011ApJ...738..162S}, and PS1 \cite{2018ApJ...859..101S}.
Furthermore, surveys like SCP, GOODS, HDFN, and CANDELS/CLASH have released high-redshift ($z > 1.0$) data \cite{2004ApJ...607..665R,2007ApJ...659...98R,2012ApJ...746...85S}. This comprehensive compilation resulted in the largest sample of SNe Ia, known as the "Pantheon+ Sample" \cite{pantheonp}, consisting of a total of 1701 light curves of 1550 spectroscopically confirmed SNe Ia spanning the redshift range $0.00122<z<2.26137$.
This compilation extends the Pantheon compilation \cite{pantheon} by enhancing the sample size and improving the treatment of systematic uncertainties associated with redshifts, peculiar velocities, photometric calibrations, and intrinsic scatter models of SNe Ia. 
To avoid potential biases arising from the considerable sensitivity of peculiar velocities at low redshifts $z < 0.008$, as depicted in Fig.~4 of Ref.~\cite{pantheonp}, the data points within the redshift range $z < 0.01$ are excluded in this analysis.

Each SNe distance ($\mu_{i}$) is compared to the predicted model distance given the measured SN/host redshift $\mu_{\mathrm{th}}(z_{i})$.
The theoretical distances are computed by 
\begin{equation}
    \mu_{\mathrm{th}}= 5\mathrm{log}(D_{L}(z_{i})/10\mathrm{pc}),
\end{equation}
where $D_{L}(z_{i})$ is the model-based luminosity distance given by
\begin{equation}
\label{eq:DL}
   D_L(z)=\frac{(1+z)}{H_0}\int^{z}_{0} \frac{d \tilde{z}}{E(\tilde{z})}.
\end{equation}
The observational distance modulus residuals is computed as
\begin{equation}
    \delta D_{i}=\mu_{i}-\mu_{\mathrm{model}}(z_{i}).
\end{equation}
Following the formalism of \cite{2011ApJS..192....1C}, in the context of the Pantheon+ dataset, the $\chi^{2}$ likelihood function can be computed by
\begin{equation}
\chi^{2}_{\mathrm{SN}}=\Delta \vec{\mu}^{T} \cdot \mathbf{C}^{-1}_{\mathrm{stat+sys}} \cdot \Delta \vec{\mu},
\end{equation}
where the covariance matrix $\mathbf{C}_{\mathrm{stat+sys}}$ is the covariance matrix including both the systematic and statistical errors, which can be found in the website\footnote{https://github.com/PantheonPlusSH0ES/DataRelease}.

The total $\chi^{2}$ combining the above datasets is
\begin{equation}
    \chi^{2}_{\mathrm{tot}}=\chi^{2}_{\mathrm{BAO}}+\chi^{2}_{\mathrm{QSO}}+\chi^{2}_{\mathrm{SN}}.
\end{equation}

\section{Cosmological Models}
\label{sec:3}
In this section, we introduce four dynamical dark energy models with different parameterizations of the equation of state (EoS) of DE, including the Chevallier-Polarski-Linder (CPL), the Barboza-Alcaniz (BA), the Jassal-Bagla-Padmanabhan (JBP), and the Feng-Shen-Li-Li (FSLL) parameterizations. 

Considering the Friedmann-Lemaitre-Robertson-Walker (FLRW) metric, with the non-flat universe filled with non-relativistic matter, dark energy, and negligible radiation, the Friedmann equation can be written as
\begin{equation}
\label{eq:Fequation}
    E^{2}(z) = H^{2}(z)/H^{2}_{0}= \Omega_{m}(1+z)^{3}+\tilde{\Omega_{\mathrm{DE}}}(z)+\Omega_{K}(1+z)^{2},
\end{equation}
where $\Omega_{m}$ and $\Omega_{K}$ represent the present values of the matter density and spatial curvature, respectively. When $\Omega_{K}=0$, it reduces to the spatially flat case. Here, the dark energy component $\tilde{\Omega_{\mathrm{DE}}}(z)$ can be expressed as
\begin{equation}
\label{eq:Ode}
    \tilde{\Omega_{\mathrm{DE}}}(z)=\Omega_{\mathrm{DE}}\times \mathrm{exp}\{ 3\int^{z}_{0} \frac{1+w(z^{\prime})}{1+z^{\prime}} dz^{\prime}\},
\end{equation}
where $\Omega_{\mathrm{DE}}$ denotes the current density of dark energy, with $\Omega_{\mathrm{DE}}=1-\Omega_{m}-\Omega_{K}$. The EoS of DE $w(z)$ is defined as $w(z)=p_{\mathrm{DE}}/\rho_{\mathrm{DE}}$, where $p_{\mathrm{DE}}$ and $\rho_{\mathrm{DE}}$ are the pressure and energy density of DE, respectively. Especially, in standard $\Lambda$CDM, $w=-1$, and the dark energy density $\tilde{\Omega_{\mathrm{DE}}}(z) = \Omega_{\mathrm{DE}}$. 

For the dynamical dark energy models, the behavior of the dark energy component varies with the redshift $z$.
For the popular CPL parametrization, also referred to as the $w_{0}w_{a}$CDM model, $w(z)$ is described by 
\begin{equation}
    w(z)=w_{0}+\frac{w_{a}z}{1+z}.
\end{equation}
The corresponding Friedmann equation can be written as
\begin{equation}
    E^{2}(z) = \Omega_{m}(1+z)^3+\Omega_{\mathrm{DE}}(1+z)^{3(1+w_{0}+w_{a})} e^{-\frac{3 w_{a} z}{1+z}} + +\Omega_{K}(1+z)^{2}.
\end{equation}

The EoS of dark energy can be generalized as $w(z)=w_{0}+\frac{z(1+z)^{n-1}}{(1+z)^{n}} w_{1}$. When setting $n=1$, it reduces the CPL parameterization, while $n=2$, it yields the BA parameterization \cite{PhysRevD.80.043521}.
In the BA model, the Friedmann equation can be given by
\begin{equation}
    E^{2}(z) = \Omega_{m}(1+z)^3+\Omega_{\mathrm{DE}}(1+z)^{3(1+w_{0})}(1+z^{2})^{\frac{3w_{1}}{2}} + +\Omega_{K}(1+z)^{2}.
\end{equation}

Another commonly used model for describing dark energy is the JBP parametrization. This parametrization expresses the equation of state for dark energy in a time-varying form, typically represented as a function of the scale factor or redshift. It refines other models by allowing more flexibility in how dark energy evolves over time, improving its ability to match observations, especially at high redshifts. In this case, $w(z)$ is expressed as
\begin{equation}
    w(z)=w_{0}+\frac{w_{a}z}{(1+z)^{2}}.
\end{equation}
And the Friedmann equation can be written as 
\begin{equation}
     E^{2}(z) = \Omega_{m}(1+z)^3+\Omega_{\mathrm{DE}}(1+z)^{3(1+ w_{1})} e^{\frac{3 w_{2} z^2}{2(1+z)^2}} + +\Omega_{K}(1+z)^{2}.
\end{equation}

Moreover, \cite{Feng:2012gf} proposed a new class of dark energy EoS parametrization, called the FSLL parametrization, which reduces to a linear form when $z\ll 1$, similar to the CPL form. In this scenario, $w(z)$ is expressed as 
\begin{equation}
    w(z)=w_{0}=\frac{z}{(1+z^{2})}w_{1}.
\end{equation}
Then, we can rewrite Eq.~\eqref{eq:Fequation} as
\begin{equation}
    E^{2}(z) = \Omega_{m}(1+z)^3+ \Omega_{\mathrm{DE}}(1+z)^{3(1+w_{0}-\frac{1}{2}w_{1})}e^{\frac{3}{2}\arctan z}(1+z^{2})^{\frac{3}{4}w_{1}} +\Omega_{K}(1+z)^{2}.
\end{equation}

\section{Results}
\label{sec:4}
In Ref.~\cite{2024arXiv240502168W}, they proposed that the LRG1 and LRG2 data points would result in the DESI BAO cosmology deviating from $w_{0}=-1$, and $w_{a}=0$ in $w_{0}w_{a}$CDM model. In our analysis, we use two BAO compilations: all DESI BAO data points (BAO) and the DESI BAO dataset removing LRG1 and LRG2 data points (BAO without LRG1\&LRG2), and the combinations of BAO, SNe and QSO datasets (BAO+SN+QSO), respectively. In this section, we present our constraint results.

\begin{figure}[htbp]
\centering
\includegraphics[width=.45\textwidth]{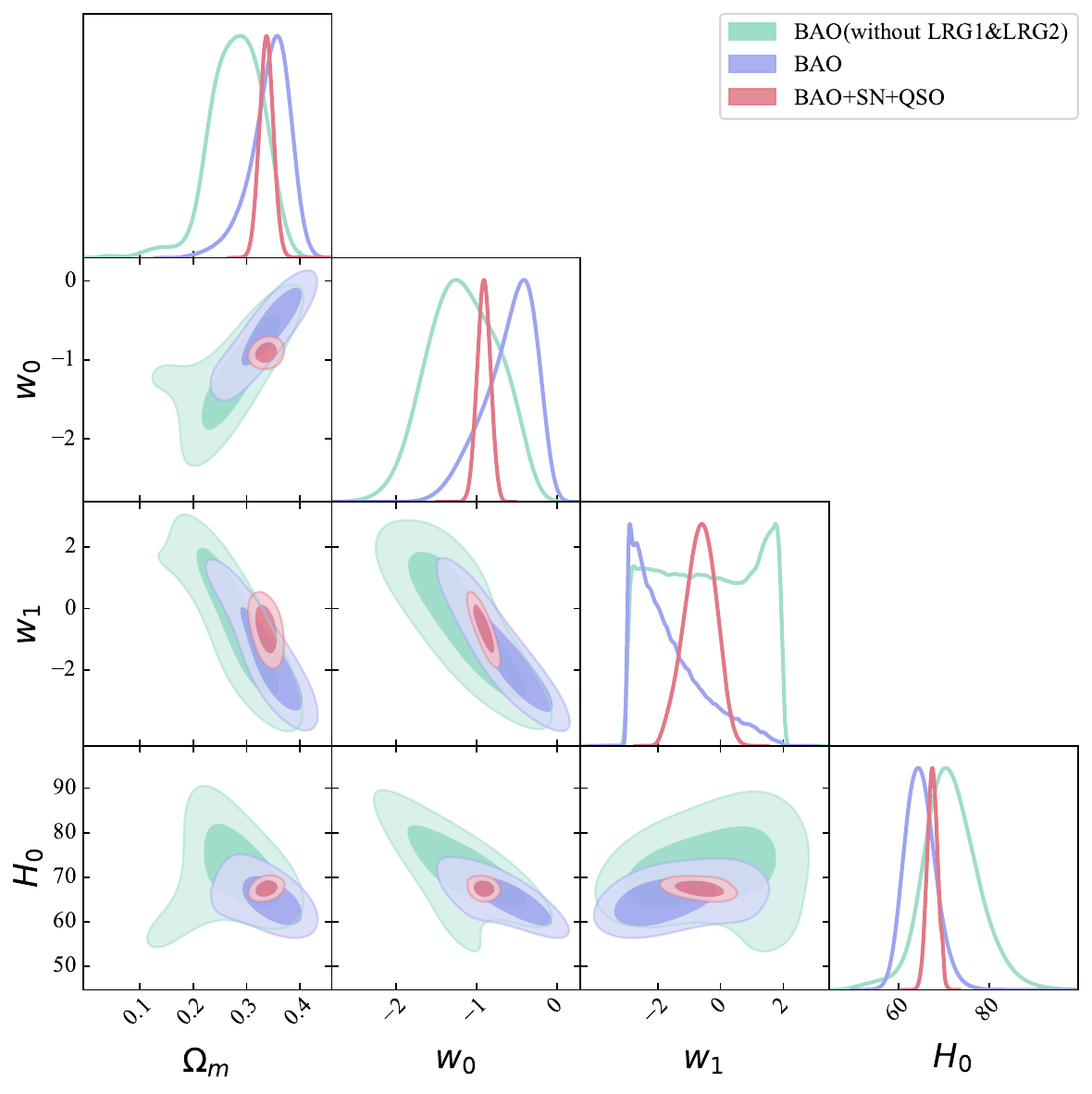}
\qquad
\includegraphics[width=.45\textwidth]{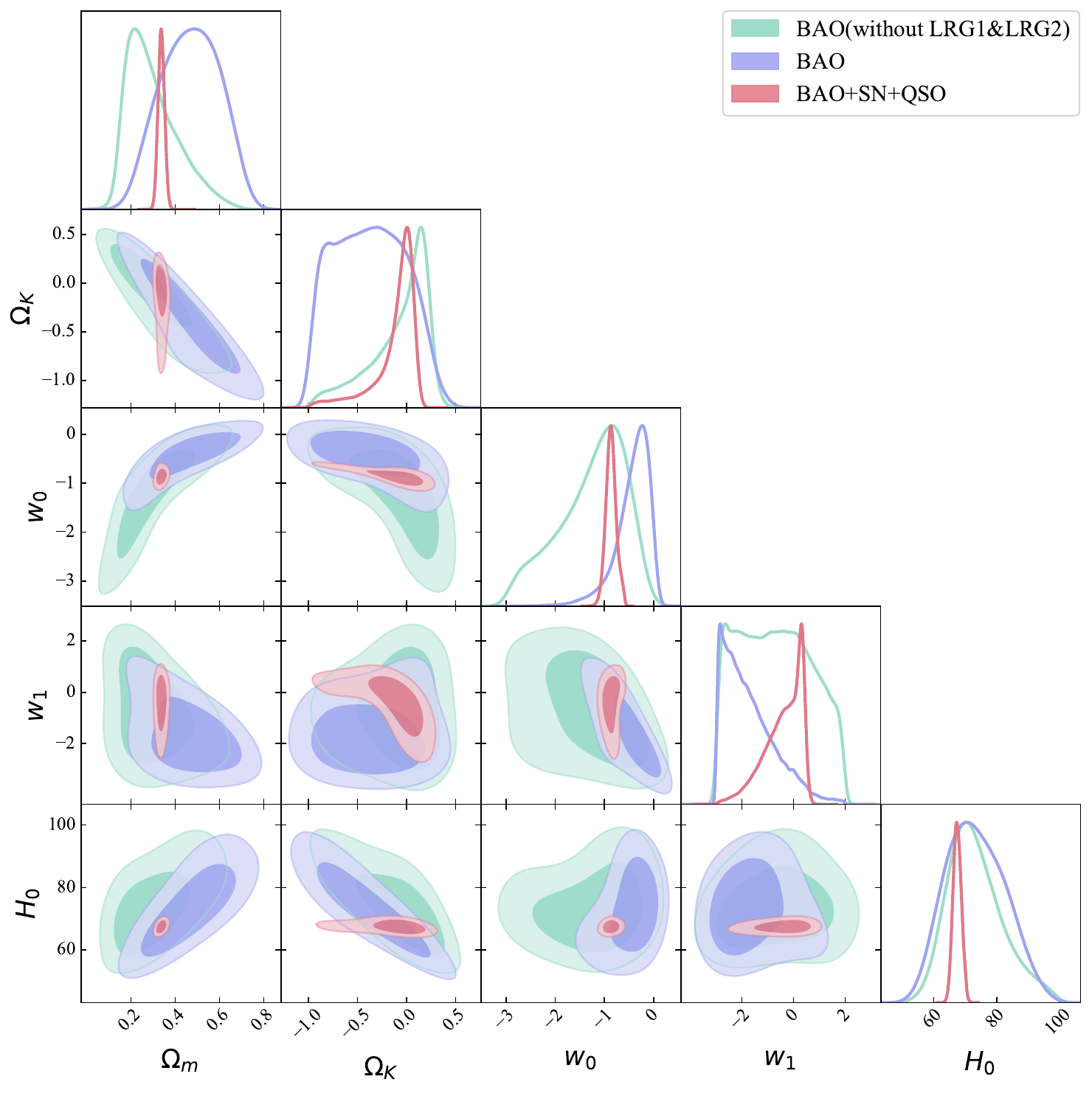}
\caption{The posterior distribution for CPL parameterization of DE derived from different datasets is illustrated, with the flat case on the left and the non-flat case on the right, respectively. \label{fig:cpl}}
\end{figure}


\paragraph{The CPL model}
As seen in the Tab.~\ref{tab:i}, the best-fit values of cosmological parameters in the flat CPL model are $\Omega_{m}=0.28\pm0.05$, $w_{0}=-1.21\pm0.37$, $w_{1}=-0.54^{+1.70}_{-1.67}$ and $H_{0}=71.24^{+6.27}_{-5.39}$ km/s/Mpc from the BAO without LRG1\&LRG2 dataset, $\Omega_{m}=0.35^{+0.03}_{-0.04}$, $w_{0}=-0.76^{+0.26}_{-0.35}$, $w_{1}=-1.50^{+1.80}_{-1.10}$, and $H_{0}=64.59^{+3.53}_{-3.18}$ km/s/Mpc from all DESI BAO data and $\Omega_{m}=0.27^{+0.15}_{-0.09}$, $w_{0}=-0.89^{+0.11}_{-0.10}$, $w_{1}=-0.89^{+0.59}_{-0.60}$ and $H_{0}=72.19^{+9.94}_{-5.40}$ km/s/Mpc from the BAO+SN+QSO dataset. 
In the non-flat CPL model, we obtain $\Omega_{m}=0.27^{+0.15}_{-0.09}$, $\Omega_{K}=0.01^{+0.18}_{-0.41}$, $w_{0}=-1.09^{+0.50}_{-0.86}$, $w_{1}=-0.64^{+1.72}_{-1.62}$ and $H_{0}=72.19^{+9.94}_{-5.40}$ km/s/Mpc from the BAO without LRG1\&LRG2 dataset, $\Omega_{m}=0.47^{+0.14}_{-0.15}$, $\Omega_{K}=-0.38\pm0.40$, $w_{0}=-0.49^{+0.24}_{-0.43}$, $w_{1}=-1.69^{+1.62}_{-0.96}$ and $H_{0}=72.80^{+10.57}_{-9.24}$ km/s/Mpc from all DESI BAO data, and $\Omega_{m}=0.34\pm0.01$, 
$\Omega_{K}=-0.05^{+0.10}_{-0.24}$, $w_{0}=-0.86^{+0.10}_{-0.11}$, $w_{1}=-0.39^{+0.91}_{-1.18}$ and $H_{0}=67.40^{+1.35}_{-1.28}$ km/s/Mpc from the combination sample.
It is clear that without the LRG1 and LRG2 data points, the constraint results do not show any significant deviation from the $\Lambda$CDM model, as they also fail to provide tight constraints on cosmological parameters.
Using the complete DESI BAO dataset, our results for the flat CPL model are consistent with those reported in Ref.~\cite{2024arXiv240403002D}. Notably, it indicates a deviation from the $\Lambda$CDM model at the 1$\sigma$ confidence level, characterized by a higher value of $w_{0}$ and a more negative $w_{1}$.
Compared to the green contours with the purple contours, it indicates that the LRG1 and LRG2 data points are beneficial for enhancing the precisions of $w_{0}$ and $w_{1}$. 
For the non-flat CPL case, introducing an additional degree of freedom in $\Omega_{K}$ slightly broadens the constraints in the $w_{0}-w_{1}$ plane. However, it does not diminish the tension with the $\Lambda$CDM model when using all DESI BAO data.
As shown in the red contours of Fig.~\ref{fig:cpl}, combining DESI BAO data with the SNe Ia and QSO observations could break the degeneracies between parameters and obtain more precise constraints on cosmological parameters, especially on $w_{0}$ and $w_{1}$.
Interestingly, the constraint results from the combined sample show slight deviations from the $\Lambda$CDM model at the 1$\sigma$ confidence levels for both flat and non-flat CPL models.
Besides, the combination sample prefers a smaller $w_{0}$ and a higher $w_{1}$, indicating an anti-correlation between $w_0$ and $w_1$.

Not only do we investigate the EoS parameters of DE, but we determine $\Omega_{m}$, $\Omega_{K}$ and $H_{0}$ simultaneously. 
The matter density parameter $\Omega_{m}$ obtained from all DESI BAO data is significantly larger than that from the BAO dataset without the LRG1 and LRG2 data points. This illustrates that the LRG1 and LRG2 data points prefer a larger $\Omega_{m}$, suggesting that including these data points could lead to different interpretations of the matter density in the universe and potentially impact the constraint results.
Regarding the Hubble constant, we find that even using the late-Universe probes, BAO+SN+QSO, the results agree well with the recent Planck 2018 result \cite{planck2018}: $H_{0}=67.4\pm0.5$ km/s/Mpc.
This alleviates the discrepancy between the Hubble constant derived from the early universe and the late universe.
For the curvature parameter $\Omega_{K}$, the determination of $\Omega_{K}$ indicates no significant deviation from flat spatial hypersurfaces. Moreover, we find that with the inclusion of LRG1 and LRG2 data points, the value of $\Omega_{K}$ shows a slight preference for a negative value.

\begin{figure}[htbp]
\centering
\includegraphics[width=.45\textwidth]{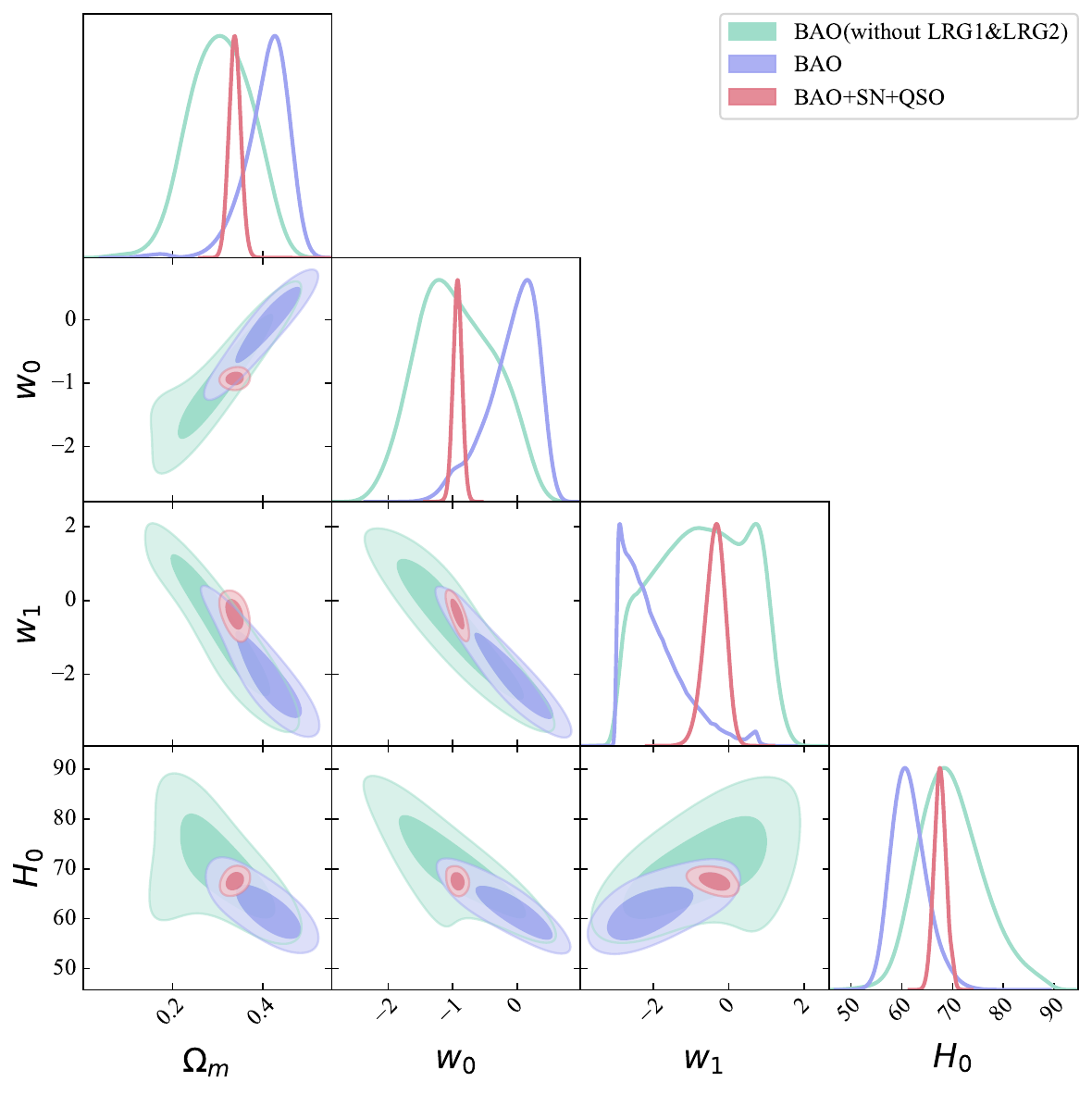}
\qquad
\includegraphics[width=.45\textwidth]{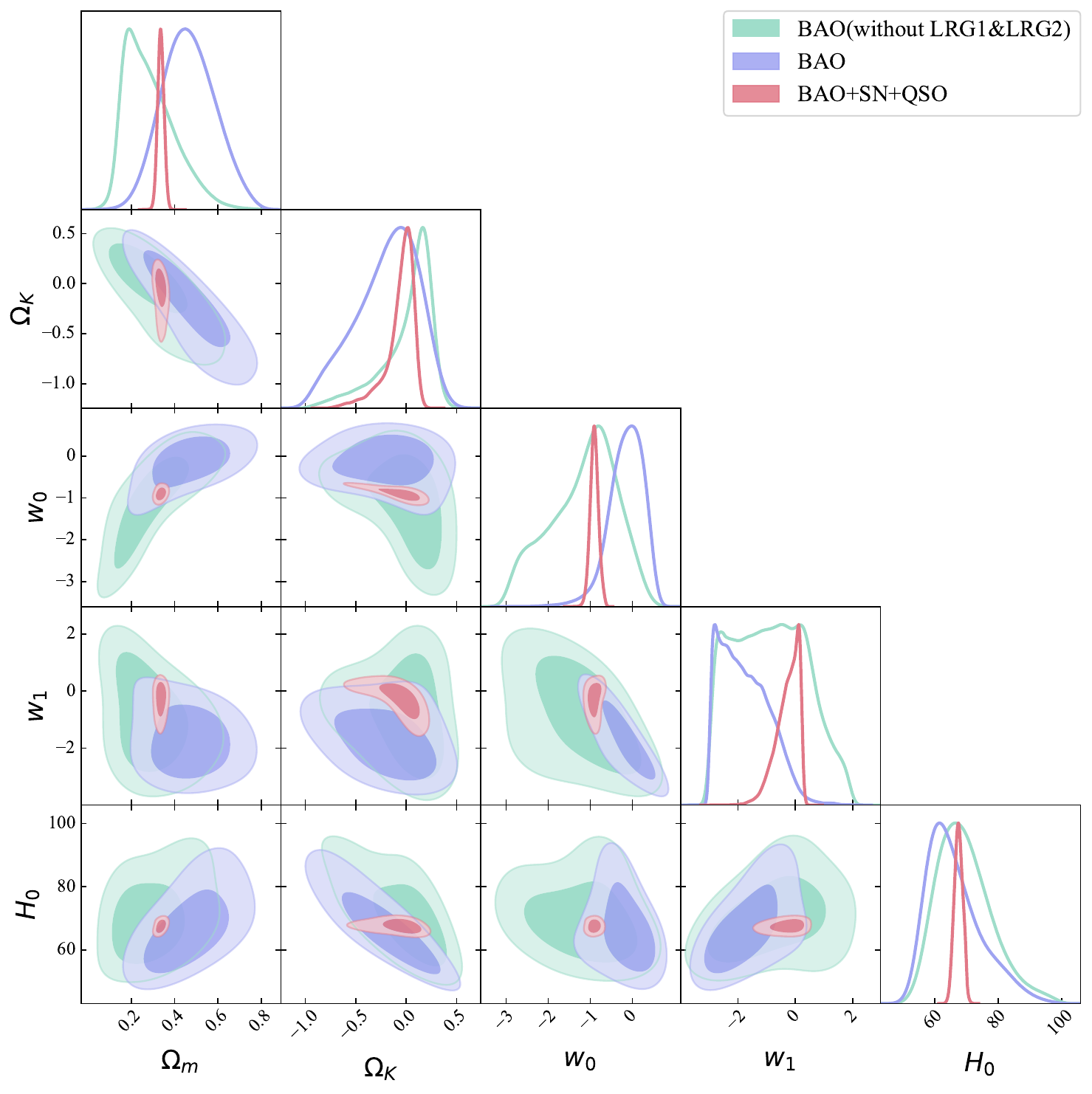}
\caption{The posterior distribution for BA parameterization of DE derived from different datasets is illustrated, with the flat case on the left and the non-flat case on the right, respectively. \label{fig:ba}}
\end{figure}

\paragraph{The BA model}
The marginalized 1$\sigma$ and 2$\sigma$ contour plots of parameters are shown in Fig.~\ref{fig:ba} and the corresponding results are listed in Tab.~\ref{tab:i}.
The best-fit values of parameters in the flat BA model are $\Omega_{m}=0.31\pm0.07$, $w_{0}=-0.98^{+0.70}_{-0.57}$, $w_{1}=-0.70^{+1.33}_{-1.39}$ and $H_{0}=69.36^{+7.00}_{-5.81}$ km/s/Mpc from the BAO without LRG1 and LRG2 data points, $\Omega_{m}=0.41^{+0.04}_{-0.05}$, $w_{0}=-0.02^{+0.31}_{-0.49}$, $w_{1}=-2.21^{+0.97}_{-0.57}$ and $H_{0}=61.21^{+3.79}_{-3.10}$ km/s/Mpc from all DESI BAO data and $\Omega_{m}=0.34\pm0.01$, $w_{0}=-0.93\pm0.07$, $w_{1}=-0.36^{+0.25}_{-0.28}$ and $H_{0}=67.53^{+1.20}_{-1.63}$ km/s/Mpc from the combination of BAO+SN+QSO dataset. 
In the non-flat BA model, we obtain $\Omega_{m}=0.27^{+0.14}_{-0.10}$, $\Omega_{K}=0.07^{+0.15}_{-0.37}$, $w_{0}=-0.98^{+0.63}_{-0.97}$, $w_{1}=-0.90^{+1.32}_{-1.42}$ and $H_{0}=68.26^{+10.19}_{-7.37}$ km/s/Mpc from the BAO without LRG1 and LRG2 data points, $\Omega_{m}=0.46^{+0.13}_{-0.12}$, $\Omega_{K}=-0.15^{+0.28}_{-0.38}$, $w_{0}=-0.11^{+0.37}_{-0.43}$, $w_{1}=-1.86^{+1.03}_{-0.81}$ and $H_{0}=64.73^{+10.42}_{-6.67}$ km/s/Mpc from all DESI BAO data, and $\Omega_{m}=0.34\pm0.01$, $\Omega_{K}=-0.03^{+0.09}_{-0.16}$, $w_{0}=-0.90\pm0.10$, $w_{1}=-0.21^{+0.83}_{-0.48}$ and $H_{0}=67.53^{+1.41}_{-1.35}$ km/s/Mpc from the combination sample.

From the above results, it seems that removing the LRG1 and LRG2 BAO data points makes the BA model more consistent with the $\Lambda$CDM model, but the errors are relatively larger.
When the LRG1 and LRG2 data points are included, the errors in $w_{0}$ and $w_{1}$ both decrease, indicating that these two data points are crucial for constraining the behavior of dark energy.
Comparing once again with a flat $\Lambda$CDM model with $\Omega_{m}=0.3$, $w_{0}=-1$, and $w_{1}=0$, the best-fits from all DESI BAO and BAO+SN+QSO samples show tensions of approximately 2$\sigma$ and 1$\sigma$ levels, respectively.
Allowing for free spatial curvature, we find that the result obtained from all DESI BAO data shows a deviation from the $\Lambda$CDM model at 1$\sigma$ level, while the result from the combination sample is consistent with the $\Lambda$CDM model within 1$\sigma$ confidence level.
Additionally, the combination of BAO, SNe Ia, and quasars data can significantly break the parameter degeneracies and improve the precision of the constraint results. Moreover, in the cases of both the flat and non-flat BA models, the combination sample could provide tight constraints on both $w_{0}$ and $w_{1}$.
In all datasets, the best-fit values agree with the ones from the other models and the correlation between parameters is the same as already obtained in the CPL model.

In the case of the BA model, the values of $\Omega_{m}$ obtained from all DESI BAO data are still larger than that obtained from the BAO data excluding the LRG1 and LRG2 data points. It indicates that these two data points prefer a larger $\Omega_{m}$. For the Hubble constant, the result from the comprehensive BAO+SN+QSO dataset is still consistent with the recent Planck 2018 result \cite{planck2018}.
Our results show that there is no significant evidence indicating the deviation of cosmic curvature in the non-flat BA model from zero (spatially flat geometry) for the determination of cosmic curvature. More specifically, the combination sample of BAO+SN+QSO favors closed geometry.

\begin{figure}[htbp]
\centering
\includegraphics[width=.45\textwidth]{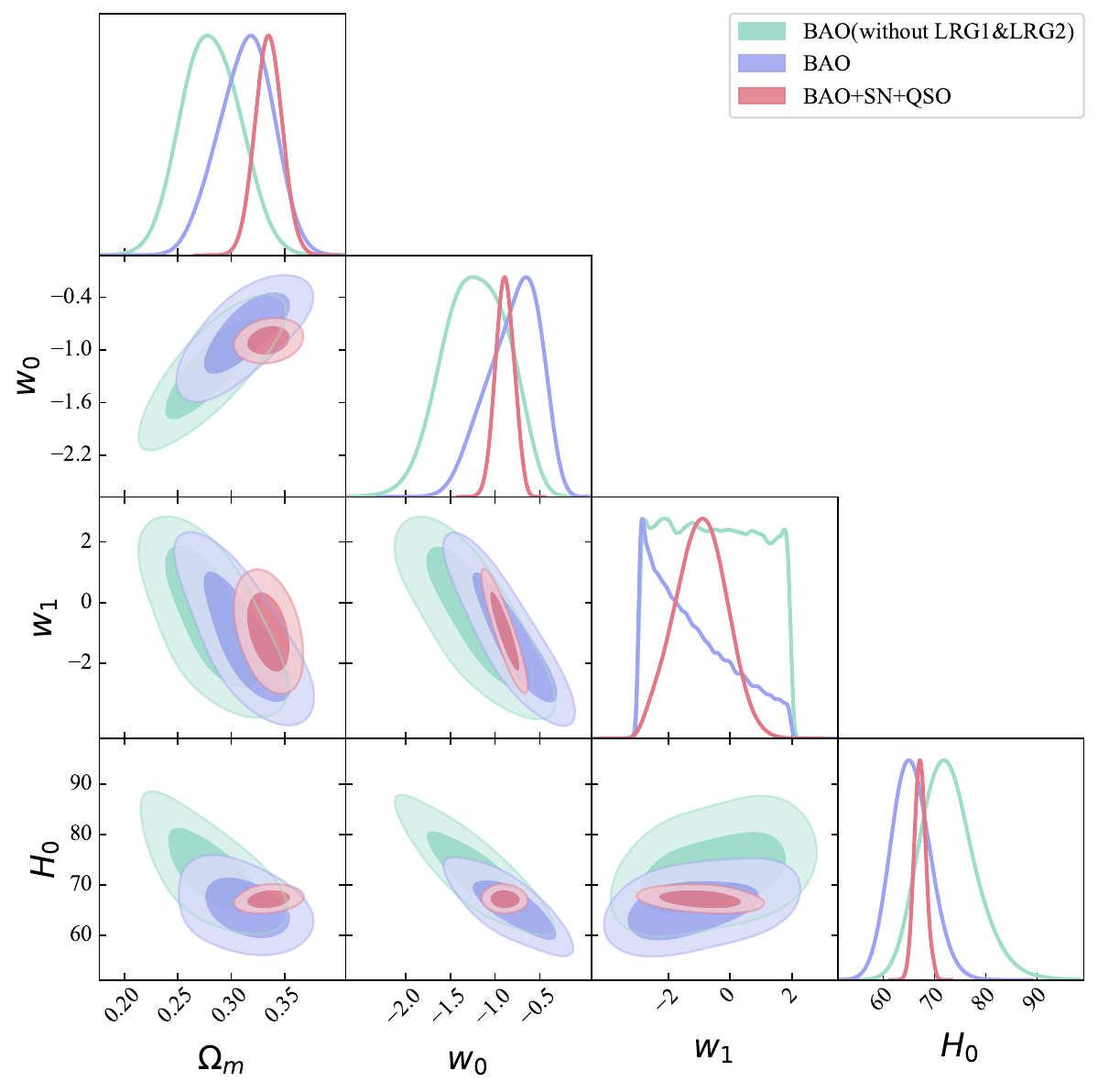}
\qquad
\includegraphics[width=.45\textwidth]{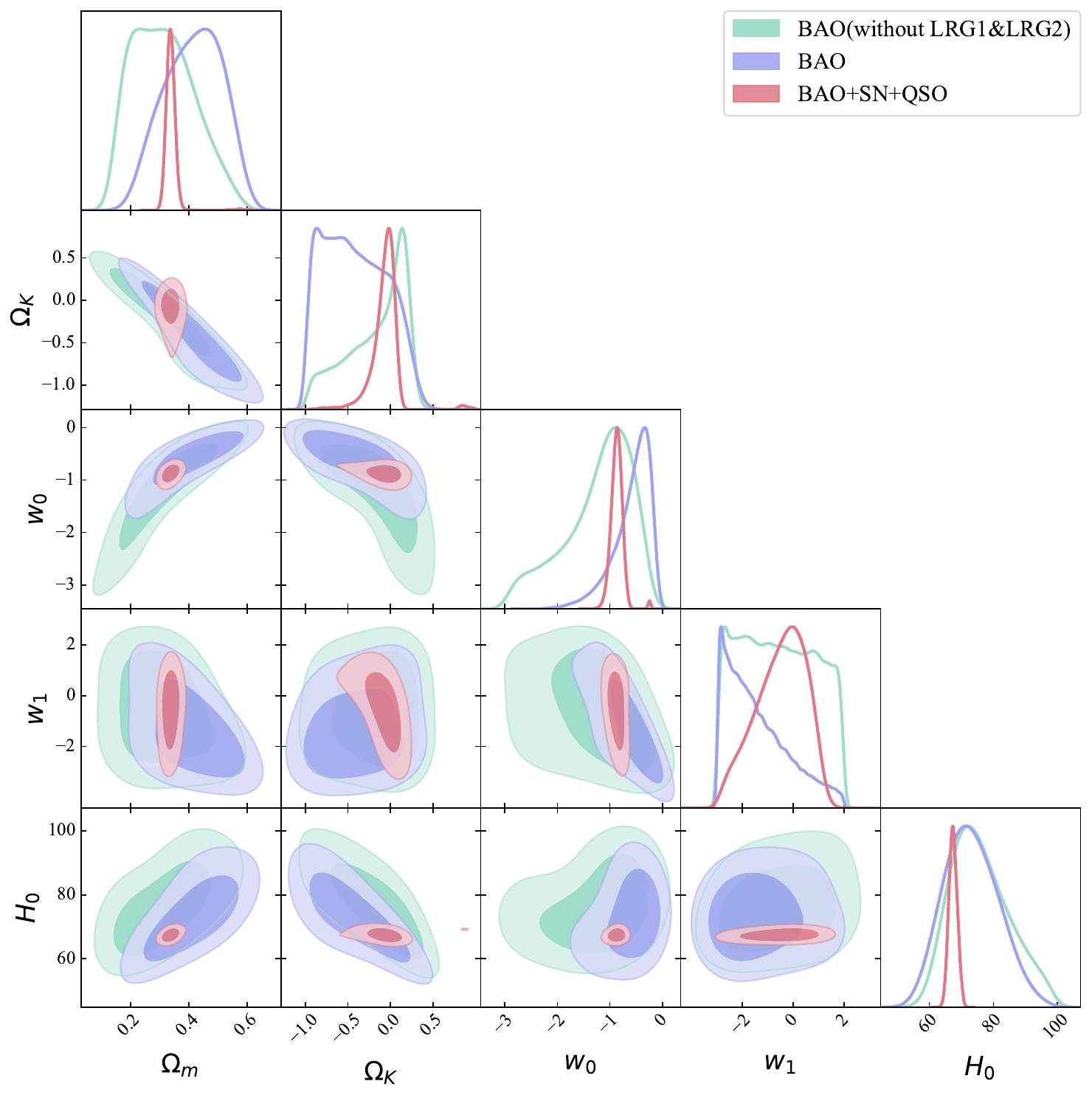}
\caption{The posterior distribution for JBP parameterization of DE derived from different datasets is illustrated, with the flat case on the left and the non-flat case on the right, respectively. \label{fig:jbp}}
\end{figure}

\paragraph{The JBP model}
The marginalized 1$\sigma$ and 2$\sigma$ contour plots of parameters are shown in Fig.~\ref{fig:jbp} and the corresponding results are listed in Tab.~\ref{tab:i}.
The best-fit values of parameters in the flat JBP model are $\Omega_{m}=0.28\pm0.03$, $w_{0}=-1.21\pm0.37$, $w_{1}=-0.54^{+1.70}_{-1.67}$ and $H_{0}=72.35^{+5.64}_{-4.93}$ km/s/Mpc from the BAO without LRG1\&LRG2, $\Omega_{m}=0.31^{+0.02}_{-0.03}$, $w_{0}=-0.76^{+0.26}_{-0.35}$, $w_{1}=-1.50^{+1.80}_{-1.10}$ and $H_{0}=65.31^{+3.96}_{-3.67}$ km/s/Mpc from all DESI BAO data and $\Omega_{m}=0.33\pm0.01$, $w_{0}=-0.89^{+0.11}_{-0.10}$, $w_{1}=-0.94^{+0.83}_{-0.87}$ and $H_{0}=67.19^{+1.14}_{-1.10}$ km/s/Mpc from the combination of BAO+SN+QSO dataset. 
In the non-flat JBP model, we obtain $\Omega_{m}=0.31^{+0.12}_{-0.11}$, $\Omega_{K}=-0.07^{+0.21}_{-0.1}$, $w_{0}=-1.09^{+0.50}_{-0.86}$, $w_{1}=-0.64^{+1.72}_{-1.62}$ and $H_{0}=74.87^{+11.01}_{-8.17}$ km/s/Mpc from the BAO without LRG1 and LRG2 data points, $\Omega_{m}=0.42^{+0.10}_{-0.11}$, $\Omega_{K}=-0.44^{+0.45}_{-0.38}$, $w_{0}=-0.49^{+0.24}_{-0.43}$, $w_{1}=-1.69^{+1.62}_{-0.96}$ and $H_{0}=72.70^{+9.40}_{-8.30}$ km/s/Mpc from all DESI BAO data, and $\Omega_{m}=0.34^{+0.02}_{-0.01}$, $\Omega_{K}=-0.05^{+0.09}_{-0.14}$, $w_{0}=-0.86^{+0.10}_{-0.11}$, $w_{1}=-0.39^{+0.91}_{-1.18}$ and $H_{0}=67.39^{+1.37}_{-1.27}$ km/s/Mpc from the combination sample.

The results show that the DESI BAO data without LRG1 and LRG2 cannot provide tight constraints on the cosmological parameters, particularly on the EoS of DE, $w_{0}$ and $w_{1}$. 
This highlights the significance of the LRG1 and LRG2 data points in constraining cosmological parameters.
Interestingly, the DESI BAO data, both alone and in combination with other cosmological probes, do not show any evidence inconsistent with the $\Lambda$CDM model when a flat JBP model is assumed.
This implies that the deviation in the case of the flat JBP model is more moderate than that of other models.
However, as for the non-flat JBP model, the results obtained from all DESI BAO data and the combination of BAO+SN+QSO data show deviations from a flat $\Lambda$CDM model at 1$\sigma$ confidence levels.
Compared to the purple and red contours in Fig.~\ref{fig:jbp}, it is evident that the combination of BAO+SN+QSO has the potential to break the degeneracy between $H_{0}$ and $w_{0}$, reducing the errors in the constraint results, while the constraint on $w_{1}$ remains loose.
The DESI BAO data prefer a higher $w_{0}$ and a more negative $w_{1}$, while the combination sample favors a smaller $w_{0}$ and a larger $w_{1}$, indicating that their correlations are similar to those in other models.

Similar to other cosmological models, all DESI BAO data prefer larger values of $\Omega_{m}$ and EoS parameter $w_{0}$, compared to the DESI BAO data excluding the LRG1 and LRG2 data points.
Moreover, the combination of the standard ruler and standard candle probes favors the $H_{0}$ value obtained from the Planck collaboration \cite{planck2018}.
For the cosmic curvature, our results show that there is no significant evidence indicating a deviation from zero (spatially flat geometry). More specifically, the combination sample of BAO+SN+QSO favors closed geometry.
Meanwhile, our results also demonstrate that zero spatial curvature is supported by the current observations and there is no significant deviation from a flat universe.


\begin{figure}[htbp]
\centering
\includegraphics[width=.45\textwidth]{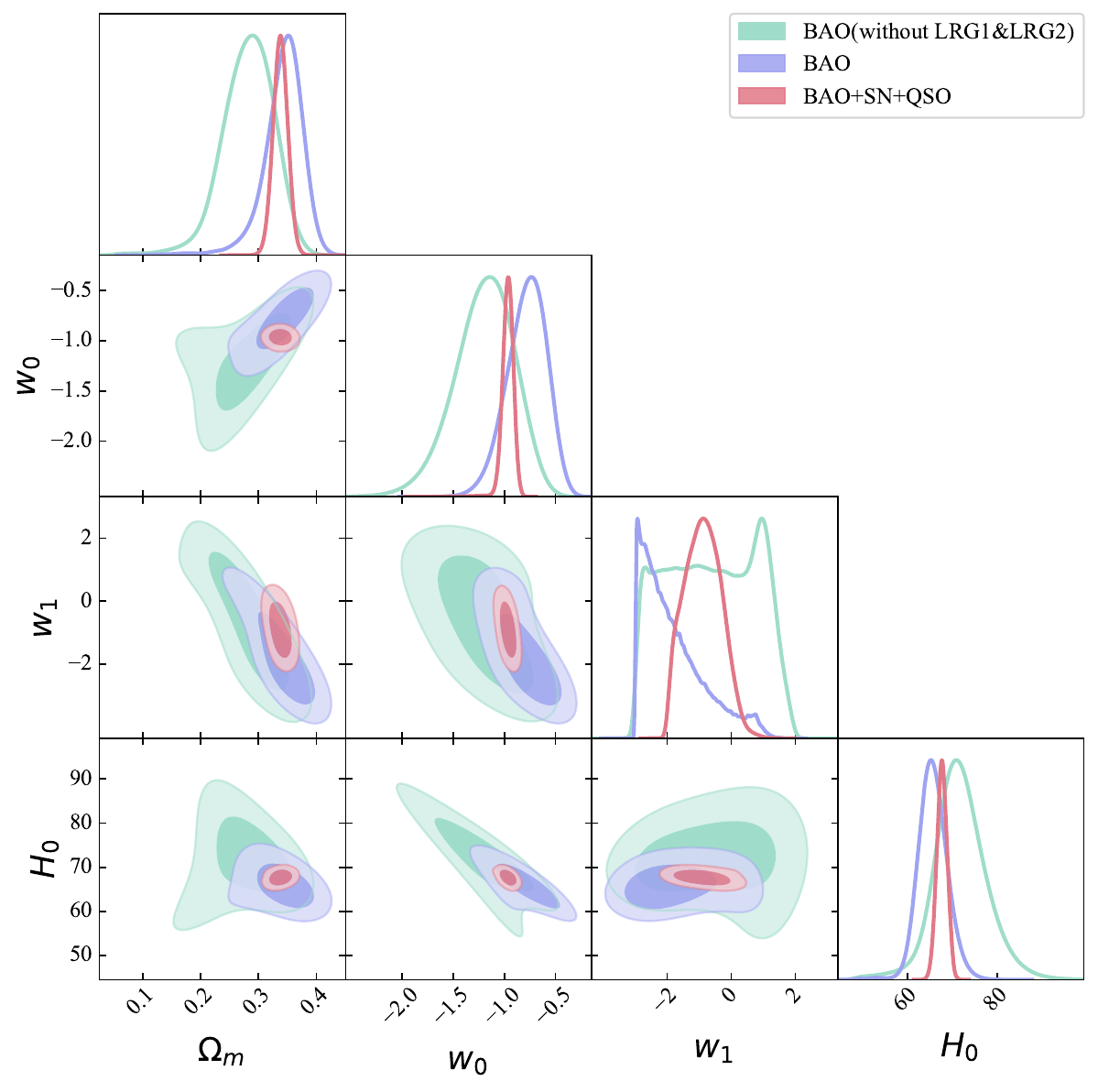}
\qquad
\includegraphics[width=.45\textwidth]{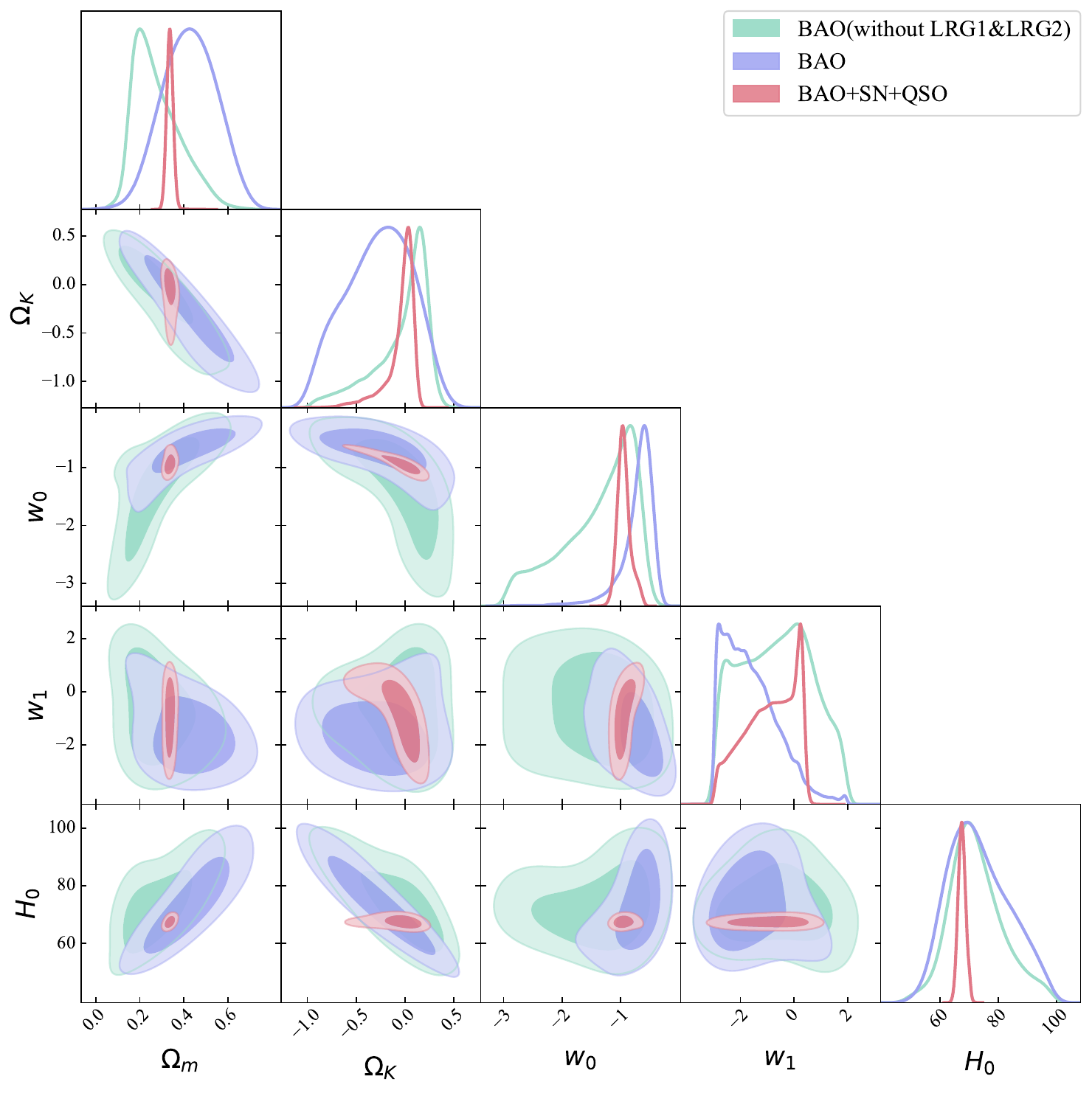}
\caption{The posterior distribution for FSLL parameterization of DE derived from different datasets is illustrated, with the flat case on the left and the non-flat case on the right, respectively. \label{fig:fsll2}}
\end{figure}

\paragraph{The FSLL model}
For the FSLL model, the marginalized 1$\sigma$ and 2$\sigma$ contour plots of parameters are shown in Fig.~\ref{fig:jbp} and the corresponding results are listed in Tab.~\ref{tab:i}.
The best-fit values of parameters in the flat FSLL model are $\Omega_{m}=0.28^{+0.04}_{-0.05}$, $w_{0}=-1.18^{+0.27}_{-0.31}$, $w_{1}=-0.65^{+1.56}_{-1.58}$ and $H_{0}=71.53^{+5.76}_{-4.97}$ km/s/Mpc from the BAO without LRG1 and LRG2 data points, $\Omega_{m}=0.35\pm0.03$, $w_{0}=-0.77^{+0.18}_{-0.21}$, $w_{1}=-2.06^{+1.25}_{-0.68}$ and $H_{0}=65.55^{+3.23}_{-2.94}$ km/s/Mpc from all DESI BAO data and $\Omega_{m}=0.33\pm0.01$, $w_{0}=-0.96^{+0.04}_{-0.05}$, $w_{1}=-0.89^{+0.59}_{-0.60}$ and $H_{0}=67.67\pm1.14$ km/s/Mpc from the combination of BAO+SN+QSO dataset. 
In the non-flat FSLL model, we obtain $\Omega_{m}=0.26^{+0.14}_{-0.08}$, $\Omega_{K}=0.05^{+0.15}_{-0.89}$, $w_{0}=-1.22^{+0.46}_{-0.88}$, $w_{1}=-0.63^{+1.38}_{-1.57}$ and $H_{0}=71.40^{+10.32}_{-7.57}$ km/s/Mpc from the BAO without LRG1 and LRG2 data points, $\Omega_{m}=0.43\pm0.13$, $\Omega_{K}=-0.25^{+0.34}_{-0.40}$, $w_{0}=-0.64^{+0.17}_{-0.29}$, $w_{1}=-1.76^{+1.19}_{-0.86}$ and $H_{0}=72.20^{+13.08}_{-9.57}$ km/s/Mpc from BAO data, and $\Omega_{m}=0.34\pm0.01$, $\Omega_{K}=0.00^{+0.07}_{-0.16}$, $w_{0}=-0.95^{+1.12}_{-0.09}$, $w_{1}=-0.81^{+0.96}_{-1.18}$ and $H_{0}=67.44^{+1.34}_{-1.27}$ km/s/Mpc from the combination sample.

For both flat and non-flat FSLL models, the DESI BAO data, excluding the LRG1 and LRG2 data points, fail to provide robust constraints on the cosmological parameters. Especially, these data points are essential for constraining the parameter $w_{0}$.
For the flat FSLL model, the results from both the DESI BAO data and the combined BAO+SN+QSO dataset are inconsistent with the $\Lambda$CDM model at 1$\sigma$ confidence levels.
As for the non-flat FSLL model, the results derived from all DESI BAO data and the combined BAO+SN+QSO samples show respectively a tension of 1$\sigma$ and a consistency at 1$\sigma$ confidence level with the flat $\Lambda$CDM model.
The correlation between each pair of parameters is the same as already seen in the CPL, BA, and JBP models.
Seen in the Fig.~\ref{fig:fsll2}, the combination of BAO+SN+QSO has the potential to break the degeneracies between $w_{0}-w_{1}$ and $\Omega_{m}-H_{0}$ planes, and yields more precise results.

Similar to other cosmological models, all DESI BAO data prefer a larger $\Omega_{m}$ compared to the DESI BAO data when excluding the LRG1 and LRG2 data points.
Although we use late-Universe probes, BAO+SN+QSO, our constraint on $H_0$ favors the value obtained by the early-Universe probe \cite{planck2018}.
Meanwhile, for the determination of cosmic curvature, our results show that zero spatial curvature (spatially flat geometry) is favored by the combination sample of BAO+SN+QSO, with no significant deviation from a flat universe.


\begin{table}
    \centering
    \resizebox{\textwidth}{!}{
    \begin{tabular}{ccccccc}
        \hline
        \hline
        Model & Data set & $\Omega_{m}$ & $\Omega_{K}$ & $w_{0}$ & $w_{1}$ & $H_{0}$(km/s/Mpc) \\
        \hline
        Flat CPL & BAO (without LRG1\&2) & $0.28\pm0.05$ & - &  $-1.17^{+0.48}_{-0.44}$ & $-0.45^{+1.77}_{-1.75}$ & $71.24^{+6.27}_{-5.39}$ \\
        & BAO & $0.35^{+0.03}_{-0.04}$ & - & $-0.54^{+0.26}_{-0.40}$ & $-1.95^{+1.45}_{-0.76}$  & $64.59^{+3.53}_{-3.18}$\\
        & BAO+SN+QSO & $0.34\pm0.01$ & - & $-0.91\pm0.08$ & $-0.64^{+0.49}_{-0.53}$ & $67.43^{+1.18}_{-1.16}$ \\
        \hline
        Flat BA &  BAO (without LRG1\&2) & $0.31\pm0.07$ & - & $-0.98^{+0.70}_{-0.57}$ & $-0.70^{+1.33}_{-1.39}$ & $69.36^{+7.00}_{-5.81}$\\
        & BAO & $0.41^{+0.04}_{-0.05}$ &  - &$-0.02^{+0.31}_{-0.49}$ & $-2.21^{+0.97}_{-0.57}$ & $61.21^{+3.79}_{-3.10}$\\
        & BAO+SN+QSO & $0.34\pm0.01$ & - & $-0.93\pm0.07$ & $-0.36^{+0.25}_{-0.28}$ & $67.53^{+1.20}_{-1.63}$ \\
        \hline
        Flat JBP & BAO (without LRG1\&2) & $0.28\pm0.03$ & - & $-1.21\pm0.37$ & $-0.54^{+1.70}_{-1.67}$ & $72.35^{+5.64}_{-4.93}$ \\
        & BAO & $0.31^{+0.02}_{-0.03}$ & - & $-0.76^{+0.26}_{-0.35}$ & $-1.50^{+1.80}_{-1.10}$ & $65.31^{+3.96}_{-3.67}$\\
        & BAO+SN+QSO & $0.33\pm0.01$ & - & $-0.89^{+0.11}_{-0.10}$ & $-0.94^{+0.83}_{-0.87}$ & $67.19^{+1.14}_{-1.10}$\\
        \hline
        Flat FSLL & BAO (without LRG1\&2) & $0.28^{+0.04}_{-0.05}$ & - & $-1.18^{+0.27}_{-0.31}$ & $-0.65^{+1.56}_{-1.58}$ & $71.53^{+5.76}_{-4.97}$ \\
        & BAO & $0.35\pm0.03$ & - & $-0.77^{+0.18}_{-0.21}$ & $-2.06^{+1.25}_{-0.68}$ & $65.55^{+3.23}_{-2.94}$\\
        & BAO+SN+QSO & $0.33\pm0.01$& - & $-0.96^{+0.04}_{-0.05}$ & $-0.89^{+0.59}_{-0.60}$ & $67.67\pm1.14$   \\
        \hline
        Non-flat CPL &  BAO (without LRG1\&2) & $0.27^{+0.15}_{-0.09}$  & $0.01^{+0.18}_{-0.41}$ & $-1.18^{+0.56}_{-0.89}$ & $-0.75^{+1.60}_{-1.54}$ & $72.19^{+9.94}_{-7.50}$ \\
        & BAO & $0.47^{+0.14}_{-0.15}$ & $-0.38\pm0.40$ & $-0.35^{+0.25}_{-0.38}$ & $-1.92^{+1.25}_{-0.77}$ & $72.80^{+10.57}_{-9.24}$ \\
        & BAO+SN+QSO & $0.34\pm0.01$ & $-0.05^{+0.10}_{-0.24}$ & $-0.87^{+0.11}_{-0.10}$ &  $-0.27^{+0.57}_{-0.90}$ & $67.40^{+1.35}_{-1.28}$\\
        \hline
        Non-flat BA &  BAO (without LRG1\&2) & $0.27^{+0.14}_{-0.10}$ & $0.07^{+0.15}_{-0.37}$ & $-0.98^{+0.63}_{-0.97}$ & $-0.90^{+1.32}_{-1.42}$ & $68.26^{+10.19}_{-7.37}$\\
        & BAO & $0.46^{+0.13}_{-0.12}$ & $-0.15^{+0.28}_{-0.38}$ & $-0.11^{+0.37}_{-0.43}$ & $-1.86^{+1.03}_{-0.81}$ & $64.73^{+10.42}_{-6.67}$ \\
        & BAO+SN+QSO & $0.34\pm0.01$ & $-0.03^{+0.09}_{-0.16}$ & $-0.90\pm0.10$ & $-0.21^{+0.33}_{-0.48}$ & $67.53^{+1.41}_{-1.35}$\\
        \hline
        Non-flat JBP &  BAO (without LRG1\&2) & $0.31^{+0.12}_{-0.11}$ & $-0.07^{+0.21}_{-0.51}$ & $-1.09^{+0.50}_{-0.86}$ & $-0.64^{+1.72}_{-1.62}$ & $74.87^{+11.01}_{-8.17}$\\
        & BAO & $0.42^{+0.10}_{-0.11}$  & $-0.44^{+0.45}_{-0.38}$ & $-0.49^{+0.24}_{-0.43}$ & $-1.69^{-1.62}_{-0.96}$ & $72.70^{+9.40}_{-8.30}$\\
        & BAO+SN+QSO & $0.34^{+0.02}_{-0.01}$ & $-0.05^{+0.09}_{-0.14}$ & $-0.86^{+0.10}_{-0.11}$ & $-0.39^{+0.91}_{-1.18}$ & $67.39^{+1.37}_{-1.27}$\\
        \hline
        Non-flat FSLL &  BAO (without LRG1\&2)  & $0.26^{+0.14}_{-0.08}$ & $0.05^{+0.15}_{-0.89}$ & $-1.22^{+0.46}_{-0.88}$ & $-0.63^{+1.38}_{-1.57}$ & $71.40^{+10.32}_{-7.57}$\\
        & BAO & $0.43\pm0.13$ & $-0.25^{+0.34}_{-0.40}$ & $-0.64^{+0.17}_{-0.29}$ & $-1.76^{+1.19}_{-0.86}$ & $72.20^{+13.08}_{-9.57}$\\
        & BAO+SN+QSO & $0.34\pm0.01$ & $-0.00^{+0.07}_{-0.16}$ & $-0.95^{+1.12}_{-0.09}$ & $-0.81^{+0.96}_{-1.18}$ & $67.44^{+1.34}_{-1.27}$\\
        \hline
        \hline
    \end{tabular} 
    }
    \caption{The best-fit values and 1$\sigma$ uncertainties for the parameters in each cosmological model.}
    \label{tab:i}
\end{table}

\begin{table*}
\centering
    \label{tab:ic}
    \begin{tabular}{llllll}
    \hline
    \hline
    \multicolumn{6}{c}{\textbf{BAO(without LRG1\&2)}} \\
    \hline
    \textbf{Models} & $\chi^{2}$ & $\mathrm{AIC}$ & $\Delta\mathrm{AIC}$ & $\mathrm{BIC}$ & $\Delta\mathrm{BIC}$ \\
    \hline
    Flat $\Lambda$CDM  & 6.67 & 10.67 & 0 &  9.89 & 0   \\
    Flat CPL & 5.48 & 13.48 & +2.81 & 10.09 &  +0.2 \\
    Flat BA & 6.49 & 14.49 & +3.82 & 11.10 & +1.21 \\
    Flat JBP & 4.75 & 12.75 & +2.08 & 9.36 & -0.53 \\
    Flat FSLL & 5.12 & 13.12 & +2.45 & 9.73 & -0.16  \\
    \hline
    Non-flat $\Lambda$CDM & 4.53 & 10.53 & 0 & 9.36 & 0 \\
    Non-flat CPL & 6.16 &  16.16 & +5.63 & 10.77 & +1.41  \\
    Non-flat BA & 6.42 & 16.42 & +5.89 &  11.03 & +1.67  \\
    Non-flat JBP & 4.78 & 14.78 & +4.25 &  9.39 & +0.03  \\
    Non-flat FSLL & 10.47 & 20.47 & +9.94&  15.08 & +5.72 \\
    \hline
    \multicolumn{6}{c}{\textbf{BAO}} \\
    \hline
    \textbf{Models} & $\chi^{2}$ & $\mathrm{AIC}$ & $\Delta\mathrm{AIC}$ &  $\mathrm{BIC}$ & $\Delta\mathrm{BIC}$ \\
    \hline
    Flat $\Lambda$CDM & 14.94 & 18.94 & 0 & 18.83 & 0 \\
    Flat CPL & 12.36 & 20.36 & +1.42 & 17.92 & -0.91  \\
    Flat BA & 9.56 & 17.56 & -1.38 & 15.12 & -3.71 \\
    Flat JBP & 14.13 & 22.13 & +3.19 & 19.69 & +0.86 \\
    Flat FSLL & 11.97 & 19.97 & +1.03 & 17.53 & -1.30\\
    \hline
    Non-flat $\Lambda$CDM & 14.84 & 20.84 & 0 & 20.68 & 0 \\
    Non-flat CPL & 15.22 & 25.22 & +4.38 & 20.77 & -0.09\\
    Non-flat BA & 11.41 & 21.41 & +0.57 & 16.96 & -3.72  \\
    Non-flat JBP & 21.75 & 31.75 & +10.91 & 27.30 & +6.62 \\
    Non-flat FSLL & 11.92 & 21.92 & +1.08 & 17.47 & -3.21 \\
    \hline
    \multicolumn{6}{c}{\textbf{BAO+SN+QSO}} \\
    \hline
    \textbf{Models} & $\chi^{2}$ &  $\mathrm{AIC}$ & $\Delta\mathrm{AIC}$ & $\mathrm{BIC}$ & $\Delta\mathrm{BIC}$ \\
    \hline
    Flat $\Lambda$CDM & 3802.99 & 3806.99 & 0 & 3819.59 & 0\\
    Flat CPL & 3803.23 & 3811.23 & +4.24 &  3819.83 & -0.24 \\
    Flat BA & 3801.48 & 3809.48 & +2.49 & 3818.08 & -1.51  \\
    Flat JBP & 3816.88 & 3824.88 & +17.89 & 3833.48 & +13.89 \\
    Flat FSLL & 3801.75 & 3809.75 & +2.76 & 3818.35 & -1.24 \\
    \hline
    Non-flat $\Lambda$CDM & 3821.40 & 3827.40 & 0 & 3846.30 & 0  \\
    Non-flat CPL & 3803.39 & 3813.39 & -14.01 & 3819.99 & -26.31 \\
    Non-flat BA & 3801.93 & 3811.93 & -15.47 & 3818.53 & -27.77  \\
    Non-flat JBP & 3804.21 & 3814.21 & -13.19 &  3820.81 & -25.49\\
    Non-flat FSLL & 3802.70 & 3812.70 & -14.70 & 3819.30 & -27.00  \\
    \hline
    \hline
\end{tabular}
    \caption{The values of AIC and BIC and their differences for different dynamical cosmological models.}
\end{table*}

\paragraph{Model Comparsion}
We also compare the dark energy models described above by evaluating them using the Akaike Information Criterion (AIC) and the Bayesian Information Criterion (BIC). They are widely used tools for model comparison that balances goodness of fit with model complexity. It helps to choose the best model by penalizing models with more parameters, thus discouraging overfitting. The formula for AIC is $\mathrm{AIC}=\chi^{2}_{\mathrm{min}}+2k$, where $k$ is the number of parameters in the model, $\chi^{2}_{\mathrm{min}}$ corresponds to the maximum likelihood of the model.
The BIC is defined as $\mathrm{BIC} = \chi^{2}_{\mathrm{min}} + k\ln \mathrm{N}$, where $k$ is the number of free parameters and $N$ denotes the number of data points. 
However, BIC is more stringent than AIC, especially for larger datasets, as it penalizes model complexity more heavily.
In particular, a model with lower $\Delta \mathrm{AIC}$ or $\Delta \mathrm{BIC}$ indicates that it is more favored by the observations.
To emphasize the differences between the $\Lambda$CDM model and other dark energy models, we set $\Delta \mathrm{AIC}$ and $\Delta \mathrm{BIC}$ for the $\Lambda$CDM model to zero, with the corresponding values presented in Table.~\ref{tab:ic}. 
According to the number of model parameters, these cosmological models can be divided into two classes: the four-parameter models including the flat CPL, JBP, BA, and FSLL models; and the five-parameter models including the non-flat CPL, JBP, BA, and FSLL models. 

The $\Lambda$CDM model performs best, as it owns the lowest AIC and BIC values using the DESI BAO data without LRG1 and LRG2 data points.
However, when the LRG1 and LRG2 data are included, the AIC and BIC values for the $\Lambda$CDM model are no longer the lowest, but the BA model gains more support. It also confirms that the LRG1 and LRG2 data points would impact the constraint results.
Moreover, a significant difference arises between the flat and non-flat models using the BAO+SN+QSO dataset. In the flat case, the $\Lambda$CDM model has the lowest AIC value, while the BA and FSLL models have lower BIC values. On the other hand, the non-flat $\Lambda$CDM model performs the worst in explaining the BAO+SN+QSO observations, as it yields the highest AIC and BIC values among all the models.
Of all the candidate models, it is obvious that the BA and FSLL models are more favored in the combination of BAO+SN+QSO.


\section{Disscussion and Conclusion}
\label{sec:5}

This study is primarily motivated by the need to examine how the parameterization of the dark energy equation of state could alleviate the deviations from the $\Lambda$CDM model using DESI BAO data. 
Therefore, we analyzed several dynamical dark energy models, such as the CPL, BA, JBP, and FSLL parameterizations, in both spatially flat and non-flat assumptions.
We fit all DESI BAO data points together with the standard candle probes, SNe Ia and QSO, since the inclusion of them is crucial to break the degeneracy of parameters and extend the Hubble diagram up to a higher redshift range.
Furthermore, we investigate the impact of the LRG1 and LRG2 data points on constraining cosmological parameters. To do so, we divide the BAO data into two catalogs in agreement with~\cite{2024arXiv240502168W}.
In the following, we briefly summarize our main results:

(i) We find that the DESI BAO data prefer a higher $w_{0}$ and a more negative $w_{1}$, which is in agreement with Ref.~\cite{2024arXiv240403002D}. In contrast, the results of the joint fit favor a more negative $w_{0}$ and a greater $w_{1}$.
Moreover, the results from both flat and non-flat cosmological models show a 1-2$\sigma$ deviation from the $\Lambda$CDM model for all DESI BAO data, with the exception of the flat JBP model. Excluding the LRG1 and LRG2 data points from the entire catalog significantly affects the results, alleviating the apparent tensions raised in the original article \cite{2024arXiv240403002D}. However, this leads to looser constraint results. Accordingly, all DESI BAO data indeed show a hint of dynamic dark energy, which is an exciting result.
The combined sample of DESI BAO, SNe Ia, and QSO observations reveals that all the cosmological models we investigated show 1$\sigma$ level inconsistencies with the $\Lambda$CDM model, except for the non-flat BA and non-flat FSLL models.
This suggests that the discrepancies with the fiducial cosmology are not only derived from DESI BAO data but also observed in other observations.
However, our analysis still does not rule out dark energy as a cosmological constant.

(ii) Our results indicate that SNe Ia and QSO observations serve as promising complementary probes to the BAO data, offering significant improvements in breaking parameter degeneracies and enhancing the precision of cosmological parameter estimation.
Moreover, these late-Universe probes yield an $H_0$ precision of approximately 2\%, nearing the level required for precision cosmology. Although using the late-Unvierse probes, we find a tension with the $H_0$ estimate from SH0ES \cite{riess} while showing a perfect agreement with that inferred from the recent Planck CMB observations \cite{planck2018}. This demonstrates that combining various late-Universe probes is essential for investigating the Hubble tension problem.

(iii) We statistically evaluate which model is more consistent with the observational data using the AIC and BIC. Excluding the LRG1 and LRG2 data points, the BAO data tend to support the $\Lambda$CDM model, while including the LRG1 and LRG2 data points, the BA and FSLL models are significantly lower than the $\Lambda$CDM and other models, especially for the BA model. The combination sample also shows preferences for the BA model and FSLL models. Concerning the ranking of competing dark energy models, the flat JBP model is substantially penalized by the current observations.

In summary, the cosmological results from DESI BAO 2024 suggest the possibility of new physics beyond the $\Lambda$CDM model. Our findings indicate that different parameterizations of the DE equation of state have minimal impact on the constraint results. Furthermore, our results also demonstrate that the LRG1 and LRG2 measurements in DESI BAO 2024 are crucial in revealing the dynamical properties of dark energy. Future precise measurements of BAO data in these redshift ranges are essential to confirm the presence of dynamic dark energy. Combining these newly late-Universe probes can significantly improve constraints on cosmological parameters, as this combination effectively breaks parameter degeneracies. Additionally, our results show a slight deviation from the $\Lambda$CDM model, which deserves further investigation to better understand the nature of dark energy. In the coming years, the full DESI dataset combined with other cosmological probes will help clarify whether the dynamic features observed in current data have a physical origin.

\acknowledgments
We sincerely thank Shuo Cao for valuable suggestions and insightful discussions. This work was supported by the National Natural Science Foundation of China (Grants Nos. 12403002 and 12305059) and the Startup Research Fund of Henan Academy of Sciences (Grants Nos.  241841221, 241841222 and 241841224). 


\bibliographystyle{JHEP}
\bibliography{biblio.bib}


\end{document}